\documentclass[9pt,conference]{IEEEtran}
\IEEEoverridecommandlockouts

\usepackage{cite}
\usepackage{amsmath,amssymb,amsfonts}
\usepackage{algorithmic}
\usepackage[linesnumbered,ruled,vlined]{algorithm2e}
\usepackage{graphicx}
\usepackage{textcomp}
\usepackage{xcolor}
\usepackage{hyperref}
\usepackage{booktabs}
\def\BibTeX{{\rm B\kern-.05em{\sc i\kern-.025em b}\kern-.08em
    T\kern-.1667em\lower.7ex\hbox{E}\kern-.125emX}}
\begin{document}

\title{SSR-Speech: Towards Stable, Safe and Robust Zero-shot Text-based Speech Editing and Synthesis}

\author{\IEEEauthorblockN{Helin Wang\IEEEauthorrefmark{2}\IEEEauthorrefmark{1}\thanks{\IEEEauthorrefmark{1} Work done during an internship at Tencent AI Lab.},
Meng Yu\IEEEauthorrefmark{3},
Jiarui Hai\IEEEauthorrefmark{2}\IEEEauthorrefmark{1},
Chen Chen\IEEEauthorrefmark{4},
Yuchen Hu\IEEEauthorrefmark{4},
Rilin Chen\IEEEauthorrefmark{5}, 
Najim Dehak\IEEEauthorrefmark{2}, and
Dong Yu\IEEEauthorrefmark{3}
}
\IEEEauthorblockA{\IEEEauthorrefmark{2}Department of Electrical and Computer Engineering, Johns Hopkins University, Baltimore, MD, USA}
\IEEEauthorblockA{\IEEEauthorrefmark{3}Tencent AI Lab, Bellevue, USA}
\IEEEauthorblockA{\IEEEauthorrefmark{4}Nanyang Technological University, Singapore}
\IEEEauthorblockA{\IEEEauthorrefmark{5}Tencent AI Lab, Beijing, China}
\IEEEauthorblockA{Email: hwang258@jhu.edu}
}

\maketitle

\begin{abstract}
In this paper, we introduce SSR-Speech, a neural codec autoregressive model designed for stable, safe, and robust zero-shot text-based speech editing and text-to-speech synthesis. 
SSR-Speech is built on a Transformer decoder and incorporates classifier-free guidance to enhance the stability of the generation process. 
A watermark Encodec is proposed to embed frame-level watermarks into the edited regions of the speech so that which parts were edited can be detected. 
In addition, the waveform reconstruction leverages the original unedited speech segments, providing superior recovery compared to the Encodec model. 
Our approach achieves state-of-the-art performance in the RealEdit speech editing task and the LibriTTS text-to-speech task, surpassing previous methods. Furthermore, SSR-Speech excels in multi-span speech editing and also demonstrates remarkable robustness to background sounds. The source code\footnote{\href{https://github.com/WangHelin1997/SSR-Speech}{https://github.com/WangHelin1997/SSR-Speech}} and demos\footnote{\href{https://wanghelin1997.github.io/SSR-Speech-Demo/}{https://wanghelin1997.github.io/SSR-Speech-Demo/}} are released.
\end{abstract}


\section{Introduction}
Nowadays, zero-shot text-based speech generation \cite{DBLP:conf/icassp/CooperLYFWCY20,DBLP:conf/icml/CasanovaWSJGP22,DBLP:conf/icml/BaiZCML022,DBLP:journals/corr/abs-2406-00654} has garnered significant attention in the speech community, particularly in areas such as speech editing (SE) and text-to-speech (TTS) synthesis. Given an unseen speaker during training, zero-shot SE focuses on modifying specific words or phrases within an utterance to align with a target transcript while preserving the unchanged portions of the original speech, and zero-shot TTS is concerned with generating the whole speech following a target transcript.
Recently proposed approaches based on large-scale speech data have significantly streamlined speech generation systems.
Non-autoregressive (NAR) models, such as SoundStorm \cite{DBLP:journals/corr/abs-2305-09636}, FluentSpeech \cite{DBLP:conf/acl/JiangYZYHRZ23}, NaturalSpeech 3 \cite{DBLP:conf/icml/JuWS0XYLLST000024}, and VoiceBox \cite{DBLP:conf/nips/LeVSKSMWMAMH23}, have been proposed for their high inference speed and stability. However, they face challenges due to their reliance on phoneme-acoustic alignment and the complexity of the training process \cite{DBLP:journals/corr/abs-2406-02328}.
In contrast, language model (LM) based autoregressive (AR) models, such as VALL-E \cite{DBLP:journals/corr/abs-2301-02111}, UniAudio \cite{DBLP:conf/icml/YangT0HLGCSZ0ZW24}, and VoiceCraft \cite{DBLP:conf/acl/Peng00MH24}, simplify the training process, but are hindered by slow and unstable inference. 
For the SE task, existing methods struggle to handle multiple spans, speech with background noise or music, and preserving the unchanged portions effectively \cite{DBLP:conf/asru/TanDYJCL21,9829827,DBLP:conf/icassp/MorrisonRJBCP21}.
In addition, since these models can easily clone a human voice, AI safety becomes a potential concern \cite{DBLP:journals/algorithms/AlmutairiE22,DBLP:conf/icassp/JuvelaW24,DBLP:conf/icml/RomanFEDFT24}.

In this work, we focus on AR models for zero-shot text-based SE and TTS, and we proposed a novel Transformer-based AR model called SSR-Speech. The main contributions of this paper are summarized as follows:\\
(i) SSR-Speech leads to stable inference. Previous AR models may generate long silence and scratching noise during generation, which produce unnatural sounding speech. Inference-only classifier-free guidance is applied to enhance the stability of the generation process.\\
(ii) The generated speech by SSR-Speech contains frame-level watermarks, which provides information about whether the audio has been produced by SSR-Speech and which part of the audio has been edited or synthesized. To achieve this, a watermark Encodec model is proposed to introduce frame-level watermarks while reconstructing the waveform.\\
(iii) SSR-Speech is robust to multi-span editing and background sounds. The training pipeline of SSR-Speech includes single-span and multi-span editing, and editing any parts of the speech, so that there is no gap between training and inference for insertion,
deletion, and substitution. In addition, the watermark encodec leverages the original unedited speech segments for the waveform reconstruction, which provides better recovery compared to the Encodec model, especially for speech with background noise or music. \\
(iv) Extensive experimental results show the effectiveness of SSR-Speech, which significantly outperforms existing methods on both the zero-shot SE and TTS tasks.

\begin{figure*}[t]
  \centering
  \includegraphics[width=15cm]{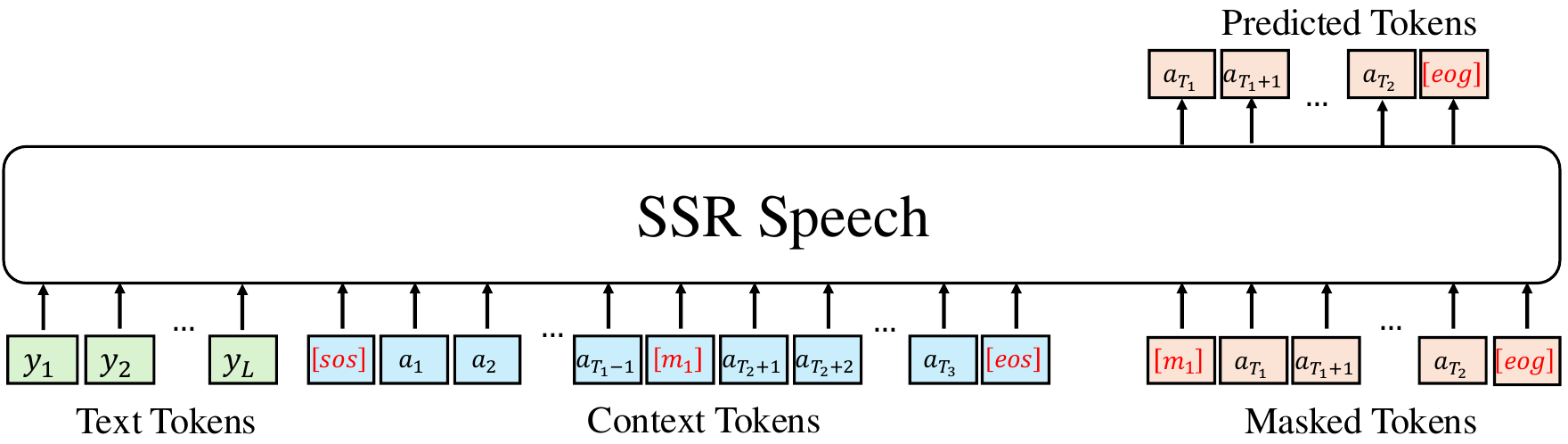}
  \caption{Diagram of SSR-Speech model. We take single-span editing as an instance, in which $\{a_{T_1},..., a_{T_2}\}$ are masked and to be predicted.
  Here, $1 \leq T_1 < T_2 \leq T_3$, where $T_3$ is the length of the audio.
  }
  \vspace{-0.1cm}
  \label{fig:1}
\end{figure*}

\section{SSR-Speech}
SSR-Speech introduces a causal Transformer decoder \cite{DBLP:conf/nips/VaswaniSPUJGKP17} that takes both text tokens and audio neural codec tokens as input and predicts the masked audio tokens in a language modeling manner.

\subsection{Modeling}
Given a speech signal, the Encodec model \cite{DBLP:journals/tmlr/DefossezCSA23} is first applied to quantize it into discrete tokens $A = \{a_1, a_2, ..., a_{T}\}$, where $T$ represents the length of the audio tokens, and each token $a_i = \{a_{i,1}, a_{i,2}, ..., a_{i,K}\}$ corresponds to $K$ codebooks of the Encodec.

As shown in Fig.~\ref{fig:1}, during training, we randomly mask $P$ continuous spans of the audio (\textit{e.g.} $P=1$ in Fig.~\ref{fig:1}). The masked tokens are concatenated with special tokens $[m_1],[m_2],...,[m_P]$, each followed by a special token $[eog]$. The unmasked tokens, also known as context tokens, are similarly concatenated with the special tokens $[m_1],[m_2],...,[m_P]$, with additional special tokens $[sos]$ and $[eos]$ at the beginning and end of the sequence, respectively. The entire set of audio tokens is then combined to form the new audio sequence $A^{\prime}=\{a^{\prime}_1, a^{\prime}_2, ..., a^{\prime}_{T^{\prime}}\}$, where $T^{\prime}$ represents the new length.

We employ a Transformer decoder to autoregressively model the masked tokens, conditioned on the speech transcript, which is embedded as a phoneme sequence $Y = \{y_1, y_2, ..., y_{L}\}$, where $L$ is the length of the phoneme tokens. At each timestep $t$ in $A^{\prime}$, the model predicts $a^{\prime}_t$ using several linear layers, conditioned on the phoneme sequence $Y$ and all preceding tokens in $A^{\prime}$ up to $a^{\prime}_t$, denoted as $X_t$.

\begin{align}
    \mathbb{P}_\theta(A^{\prime} \mid Y)= \prod_t \mathbb{P}_\theta\left(a^{\prime}_t \mid Y, X_t\right)
\end{align}
where $\theta$ denote the parameters of the model.
The training loss is defined as the negative log likelihood:
\begin{align}
\mathcal{L}(\theta)=-\log \mathbb{P}_\theta(A^{\prime} \mid Y)
\end{align}

Following \cite{DBLP:conf/acl/Peng00MH24}, we implement causal masking, delayed stacking, and apply larger weights to the first codebook than the later ones. Unlike \cite{DBLP:conf/acl/Peng00MH24}, we calculate the prediction loss only on the masked tokens, excluding special tokens, rather than on all tokens. This approach yields similar results while reducing training costs in our experiments. Additionally, we mask all regions of the audio, including the beginning and end of the speech, to better align with real-world applications. To further enhance TTS training, we also enforce speech continuation \cite{DBLP:conf/icassp/MaitiPCJC024} by consistently masking the end of the speech with a certain probability.

\subsection{Inference}
For the SE task,
we compare the original transcript with the target transcript to identify the words that need to be masked. Using word-level forced alignment\footnote{https://github.com/m-bain/whisperX} of the original transcript, we locate the corresponding masked spans of audio tokens. The phoneme tokens from the target transcript and the unmasked audio tokens are then concatenated and fed into the SSR-Speech model to autoregressively predict new audio tokens. Similar to \cite{DBLP:conf/acl/Peng00MH24}, when editing speech, the neighboring words surrounding the span to be edited also need to be slightly adjusted to accurately model co-articulation effects. Thus, we introduce a small margin hyperparameter $\alpha$, extending the length of the masked span by $\alpha$ on both the left and right sides.

For the TTS task,
the transcript of a voice prompt is combined with the target transcript of the speech to be generated. Along with the audio tokens of the voice prompt, these inputs are fed into the SSR-Speech model.

Due to the stochastic nature of autoregressive generation, the model occasionally produces excessively long silences or drags out certain sounds, resulting in unnatural-sounding speech. Previous methods address this issue by generating multiple output utterances using different random seeds and discarding the longest ones, but this approach is unstable and time-consuming. In this paper, we propose to use classifier-free guidance (CFG) \cite{ho2021classifierfree} to resolve this problem.

CFG is particularly useful in controlling the trade-off between fidelity to the input and the quality or creativity of the output for diffusion models, also used in AR generation \cite{DBLP:conf/iclr/KreukSPSDCPTA23}.
Existing methods involves training the model in two modes: conditioned and unconditioned, learning both how to generate general outputs and how to generate outputs that match a specific conditioning input.
During inference, CFG guides the model by combining the outputs from the conditioned and unconditioned modes. 
In our initial experiments, we found that traditional CFG cannot solve the dead loop of AR models well, and it may make the training unstable at the beginning.
To address this issue, we propose to use inference-only CFG that we do not need unconditioned training. More specifically, at inference time we use a random text sequence as the unconditional input, and
sample from a distribution obtained by a linear combination of the conditional and unconditional probabilities.
Formally we sample from,
\begin{align}
\gamma \mathbb{P}_\theta(A^{\prime} \mid Y) + (1-\gamma)\mathbb{P}_\theta(A^{\prime} \mid Y^{\prime})
\end{align}
where $\gamma$ is the guidance scale and $Y^{\prime}$ is a random phoneme sequence with the same length of $Y$ to enable GPU parallel processing.

Furthermore, we observed that CFG often generates speech at an accelerated pace due to the excessive removal of silence tokens during processing. To address this issue, we propose to utilize CFG with a stride of $\beta$ during inference, where $\beta$ serves as a hyperparameter.

\begin{algorithm}[h]
\SetAlgoLined
\SetKwFunction{FNextToken}{NextToken}
\SetKwProg{Fn}{Function}{:}{}
\SetKwFunction{FConcat}{concat}
\SetKwInOut{Input}{Input}
\SetKwInOut{Output}{Output}
\SetKwFunction{FSample}{Sample}

\Fn{Inference-only CFG}{$\mathcal{M}$, $A^{\prime}$, $Y$, $Y^{\prime}$, $\beta$, $\gamma$, $[sog]$, $[eog]$}{
    \Input{Model $\mathcal{M}$, initial audio sequence $A^{\prime}$, target phoneme sequence $Y$, random phoneme sequence $Y^{\prime}$, stride $\beta$, guidance scale $\gamma$, start token $[sog]$, end token $[eog]$}
    \Output{Generated sequence $S$}
    $S \leftarrow \FConcat{$A^{\prime}$, $[sog]$}$\;
    
    $t \leftarrow 1$\;
    
    \While{True}{
        \If{$t \mod \beta = 0$}{
            $s_t \sim \gamma \mathbb{P}_\theta(S \mid Y) + (1-\gamma)\mathbb{P}_\theta(S \mid Y^{\prime})$\;
        }
        \Else{
            $s_t \sim \mathbb{P}_\theta(S \mid Y)$\;
        }
        $S \leftarrow \FConcat{$S$, $s_t$}$\;
        
        \If{$s_t == [eog]$}{
            \textbf{break}\;
        }
        
        $t = t + 1$
    }
    
    \Return{$S$}\;
}
\caption{Inference-only Classifier-free Guidance}
\end{algorithm}
\vspace{-0.2cm}

\section{Watermark Encodec}
\begin{figure}[t]
  \centering
  \includegraphics[width=9cm]{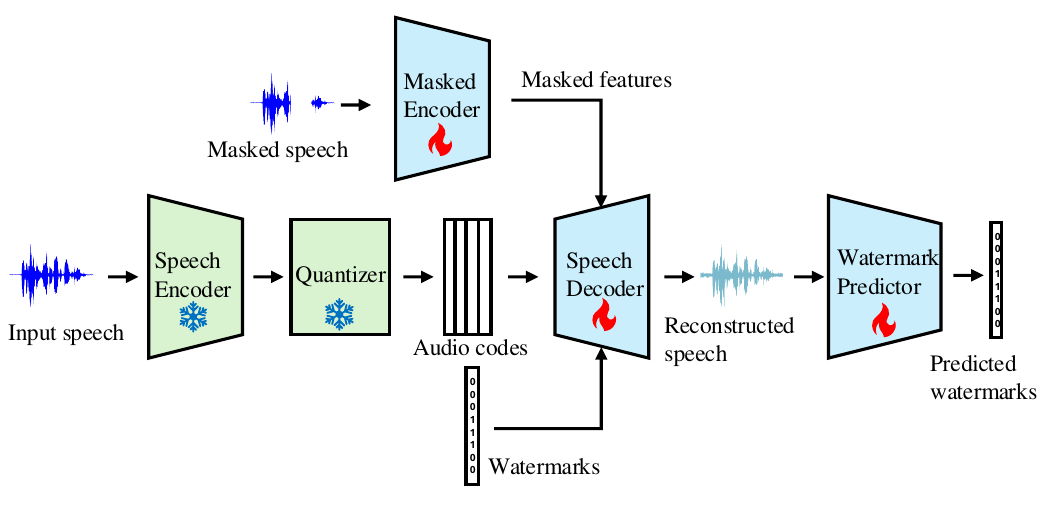}
  \caption{Diagram of the watermark Encodec model. During training, the parameters of the speech encoder and quantizer are kept frozen, while we update the speech decoder, masked encoder, and watermark predictor.}
  \label{fig:2}
\end{figure}

In this section, we introduce the watermark Encodec, a neural codec model specifically designed for the SE task, capable of watermarking the generated audio and better preserving the unedited regions. Watermark Encodec can also be applied to the TTS task. As shown in Fig.~\ref{fig:2}, the watermark Encodec consists of a speech encoder, a quantizer, a speech decoder, a masked encoder, and a watermark predictor.


\subsection{Watermarking (WM)}
The speech encoder shares the same network architecture as the encoder in Encodec. The watermark predictor also adopts the same architecture as the Encodec encoder, with the addition of a final linear layer for binary classification. We first pretrain the Encodec\footnote{https://github.com/facebookresearch/audiocraft} model and initialize the parameters of the speech encoder and watermark predictor using the pretrained Encodec encoder parameters. The quantizer is identical to the Encodec quantizer, with the same parameters copied over.

The speech decoder, which takes watermarks and audio codes as input, reconstructs the speech and shares the same architecture as the Encodec decoder. The only difference is extra linear layers to project the combined features into the same dimension as the audio features. We also initialize the speech decoder's parameters from the Encodec model. During training, the speech encoder and quantizer are frozen. The watermark is a binary sequence with the same length as the audio frames output by the speech encoder, where masked frames are marked with a value of 1, and unmasked frames are marked with 0.
An embedding layer is applied to the watermarks to obtain the watermark features.

\subsection{Context-aware Decoding (CD)}
Encodec reconstructs the waveform using audio codes. However, for the SE task, it's crucial that the unedited spans of the speech remain unchanged. To better utilize the information from these unedited spans during decoding, we propose a context-aware decoding method, which uses the original unedited waveform as an additional input to the watermark Encodec decoder.

Specifically, we mask the edited segments of the original waveform with silence clips and then use a masked encoder to extract the features from this masked waveform. The masked encoder shares the same architecture as the Encodec encoder and is initialized with parameters from Encodec. Consequently, the input to the speech decoder includes the audio codes, the watermarks, and the masked features.

Moreover, we found that using skip connections \cite{DBLP:conf/miccai/RonnebergerFB15} improves reconstruction quality and accelerates model convergence. Therefore, we fuse multi-scale features between each block, following the approach used in UNet \cite{DBLP:conf/miccai/RonnebergerFB15}.

\begin{table*}[t]
  \caption{Results for speech editing on ReadEdit. $\star$ runs inference 10 times with different margin parameters \cite{DBLP:conf/acl/Peng00MH24}. Others run once.}
  \label{tab:result1}
  \footnotesize
  \centering
  \begin{tabular}{ccccccc}
    \toprule
    \textbf{Method}&\textbf{WER $\downarrow$}&\textbf{MOSNet $\uparrow$}&\textbf{MOSNet-M $\uparrow$}&\textbf{MOSNet-N $\uparrow$}&\textbf{SpkSIM $\uparrow$}&\textbf{MOS $\uparrow$}\\
    \midrule
    GrondTruth  & $0.047\pm0.005$ & $3.700\pm0.013$& $3.692\pm0.014$& $3.520\pm0.015$ & - & $4.139\pm0.060$\\
    \midrule
    FluentSpeech & $0.052\pm0.005$ & $3.250\pm0.012$& $3.302\pm0.012$& $3.148\pm0.012$ & $0.882\pm0.004$& -\\
    VoiceCraft  & $0.069\pm0.006$ & $3.460\pm0.014$& $3.355\pm0.015$& $3.320\pm0.011$ & $0.903\pm0.003$ & $3.973\pm0.059$\\
    VoiceCraft$\star$  & $0.054\pm0.006$ & $3.655\pm0.013$& $3.596\pm0.012$ & $3.478\pm0.012$ & $0.914\pm0.003$ & -\\
    \midrule
    SSR-Speech & $\boldsymbol{0.048}\pm0.008$ & $3.707\pm0.012$& $\boldsymbol{3.694}\pm0.013$& $\boldsymbol{3.501}\pm0.012$ & $0.929\pm0.003$& $\boldsymbol{4.121}\pm0.046$ \\
    w/o WM   & $\boldsymbol{0.048}\pm0.008$ & $\boldsymbol{3.709}\pm0.012$& $\boldsymbol{3.694}\pm0.013$& $3.500\pm0.012$ & $\boldsymbol{0.930}\pm0.003$& - \\
    w/o CD   & $0.052\pm0.005$ & $3.681\pm0.013$& $3.688\pm0.014$& $3.455\pm0.012$ & $0.922\pm0.003$& - \\
    w/o CFG & $0.058\pm0.006$ & $3.578\pm0.015$& $3.499\pm0.014$& $3.398\pm0.014$ & $0.914\pm0.003$& - \\
    \bottomrule
    
  \end{tabular}
\end{table*}

\begin{table*}[t]
  \caption{Results for TTS on LibriTTS. All models run inference once.}
  \label{tab:result2}
  \footnotesize
  \centering
  \begin{tabular}{cccccc}
    \toprule
    \textbf{Method}&\textbf{WER $\downarrow$}&\textbf{MOSNet $\uparrow$}&\textbf{SpkSIM $\uparrow$}&\textbf{MOS $\uparrow$}&\textbf{SMOS $\uparrow$}\\
    \midrule
    GrondTruth  & $0.036\pm0.009$ & $3.795\pm0.012$ & $0.959\pm0.002$& $3.792\pm0.085$& - \\
    \midrule
    VALL-E  & $0.100\pm0.006$ & $3.171\pm0.012$ & $0.667\pm0.010$&-&-\\
    Pheme & $0.075\pm0.012$ & $3.214\pm0.015$ & $\boldsymbol{0.922}\pm0.006$&-&-\\
    VoiceCraft  & $0.066\pm0.014$ & $3.530\pm0.022$ & $0.912\pm0.006$& $3.346\pm0.121$& $4.014\pm0.092$\\
    SSR-Speech & $\boldsymbol{0.062}\pm0.018$ & $\boldsymbol{3.744}\pm0.018$ & $0.914\pm0.006$& $\boldsymbol{3.575}\pm0.097$& $\boldsymbol{4.106}\pm0.101$\\
    \bottomrule
  \end{tabular}
\end{table*}

\section{Experiments}
\subsection{Data}
For the SSR-Speech model, we use the Gigaspeech XL set \cite{DBLP:conf/interspeech/ChenCWDZWSPTZJK21} as the training data, which contains 10k hours of audio at a 16kHz sampling rate. Audio files shorter than 2 seconds or longer than 15 seconds are excluded. The Encodec model and the Watermark Encodec model are trained on the Gigaspeech M set, comprising 1k hours of audio data.

For the zero-shot SE task, we use the RealEdit dataset \cite{DBLP:conf/acl/Peng00MH24}, which includes 310 manually-crafted speech editing examples. For the zero-shot TTS task, we construct a dataset of 500 prompt-transcript pairs from the LibriTTS test set \cite{DBLP:conf/interspeech/ZenDCZWJCW19}. The voice prompts are between 2.5 and 4 seconds in length, and the target transcripts are randomly selected from different utterances across the entire LibriTTS test set.
\subsection{Setups}
Following \cite{DBLP:conf/acl/Peng00MH24}, both the Encodec and Watermark Encodec models use 4 RVQ codebooks, each with a vocabulary size of 2048. They operate with a stride of 320 samples, resulting in a codec framerate of 50Hz for audio recorded at a 16kHz sampling rate. The base dimension is 64, doubling at each of the 5 convolutional layers in the encoder.
The number of spans to be masked, denoted as $P$, is uniformly sampled between 1 and 3. here
can be intervals between different spans. The maximum masking length is set to $90\%$ of the original audio length. During training, we apply a probability of 0.5 to enhance TTS training. Text transcripts are phonemized using an IPA phoneset toolkit\footnote{https://github.com/bootphon/phonemizer} \cite{DBLP:journals/jossw/BernardT21a}.

The SSR-Speech model has the same architecture as VoiceCraft, which consists of 16 Transformer layers with hidden size of 2048 and 12 attention heads. The output of the final layer is passed through four separate 2-layer MLP modules to generate prediction logits. Following VoiceCraft, we employ the ScaledAdam optimizer and Eden scheduler \cite{DBLP:journals/corr/abs-2401-07333}, with a base learning rate of 0.05, a batch size of 400k frames, and a total of 50k training steps with gradient accumulation. The weighting hyperparameters for the 4 codebooks are set to $(5, 1, 0.5, 0.1)$. The SSR-Speech model has 830M parameters and was trained on 8 NVIDIA V100 GPUs for 2 weeks.

For inference, we use nucleus sampling \cite{DBLP:conf/iclr/HoltzmanBDFC20} with $p = 0.8$ and a temperature of 1. The extended masked span $\alpha$ is set to $0.12$ seconds. Based on initial experiments, we determined the optimal value for the hyperparameter $\gamma$ to be $1.5$ and $\beta$ to be $5$.

\subsection{Baselines}
For the SE task, we compare SSR-Speech with the state-of-the-art model VoiceCraft and a diffusion-based model FluentSpeech.
For the TTS task, we compare SSR-Speech with state-of-the-art autoregressive models,
including VALL-E, Phonme \cite{DBLP:journals/corr/abs-2401-02839} and VoiceCraft.
For a fair comparison, we take the original VoiceCraft\footnote{https://github.com/jasonppy/VoiceCraft} and Phonme\footnote{https://github.com/PolyAI-LDN/pheme} that were trained on the GigaSpeech dataset, and train FluentSpeech\footnote{https://github.com/Zain-Jiang/Speech-Editing-Toolkit} and VALL-E\footnote{https://github.com/open-mmlab/Amphion/tree/main/models/tts/valle} on the GigaSpeech dataset.

\subsection{Metrics}
Following prior studies, we use WER and SIM as objective evaluation metrics, calculated with pre-trained Whisper-medium.en\footnote{https://huggingface.co/openai/whisper-medium.en} \cite{DBLP:conf/icml/RadfordKXBMS23} and WavLM-TDCNN\footnote{https://huggingface.co/microsoft/wavlm-base-plus-sv} \cite{DBLP:journals/jstsp/ChenWCWLCLKYXWZ22} models for speech and speaker recognition, respectively. Additionally, we employ MOSNet\footnote{https://github.com/nii-yamagishilab/mos-finetune-ssl} \cite{DBLP:conf/icassp/CooperHTY22} to estimate an objective MOS score for reference.

For the SE task, we also report MOS estimates for noisy test samples with estimated SNRs below $20$ dB using the Brouhaha\footnote{https://github.com/marianne-m/brouhaha-vad} \cite{DBLP:conf/asru/LavechinMTBCRBCDB23} (MOSNet-N), which includes $36$ samples in the RealEdit dataset.
In addition, we test MOS estimates for multi-span editing in the RealEdit dataset (MOSNet-M), in which we have $40$ samples with $2$-span editing in total.

For subjective evaluation, we invited $20$ listeners to conduct MOS and Similar MOS (SMOS) assessments, using $60$ randomly selected samples from the RealEdit and LibriTTS test sets.
We report all these metrics with $95\%$ confidence interval.

\subsection{Results}
Table \ref{tab:result1} presents the results of the speech editing evaluations on RealEdit. 
SSR-Speech outperforms the baselines across all metrics. From the ablation studies, we observed that inference-only CFG significantly contributes to the performance improvement, effectively resolving the long silence issue in the AR model. By reducing the probabilities of undesired tokens, inference-only CFG ensures more stable and
natural generation. From the MOSNet-M results, this advantage is especially pronounced in multi-span editing. Comparing with VoiceCraft that would run inference for 10 times using different margin parameters,
our proposed SSR-Speech is able to inference once and obtain a stable result.

Context-aware decoding (CD) also enhances performance, particularly in speech with background sounds, as the unedited spans provide additional context for the model.
Therefore, SSR-Speech with CD gets much better MOSNet-N results than the others,
which bypasses the quantizer for unedited spans, preserving the unchanged parts of the audio more effectively.
While watermarking (WM) does not impact performance, it adds frame-level watermarks to the synthesized audio, increasing the model's security.
Both CFG
and watermark encoding independently contribute to improved performance in speech editing, while their combination achieves the best results.

Consistent with previous work \cite{DBLP:journals/corr/abs-2406-04840,DBLP:conf/icml/RomanFEDFT24}, we found that our watermark detector achieves a binary classification accuracy of $99.9\%$ for detecting watermarks, demonstrating a strong ability to distinguish which parts of an audio sample have been edited by SSR-Speech.

Table~\ref{tab:result2} reports the results of TTS evaluations on LibriTTS.
SSR-Speech demonstrates strong performance across multiple metrics, indicating that it produces high-quality, natural-sounding speech with excellent speaker similarity. 
Compared to VoiceCraft, we attribute the performance improvement primarily to the TTS-enhanced training and the inference-only CFG in SSR-Speech.

\section{Conclusions}
In this paper, we proposed SSR-Speech, a stable, safe, and robust zero-shot text-based SE and TTS model. SSR-Speech is a neural codec language model, which ensures strong inference stability and robustness for multi-span editing and background noise. We also introduced a watermark Encodec model to embed watermarks in the generated speech. Experiments on RealEdit and LibriTTS show that  SSR-Speech could achieve the state-of-the-art results.
Furthermore, we experimented with training SSR-Speech on Mandarin data, and the model demonstrated solid performance in the Mandarin language.
To facilitate speech
generation and AI safety research, we fully open
source our model weights.
For future works, we plan to: (i) explore more advanced neural codec models, (ii) expand to other generation tasks such as instructive TTS and voice conversion, (iii) scale up training on larger datasets and more languages, and (iv) investigate editing the prosody of speech.

{

\bibliographystyle{IEEEtran}
\bibliography{IEEEexample}
}

\end{document}